\documentclass[article,prd,epsf,amsmath,amssymb,nofootinbib,10pt]{revtex4}
\usepackage{amssymb}
\usepackage{bm}
\usepackage{latexsym}
\usepackage{amsfonts,amssymb}
\usepackage{graphicx,epsfig}
\usepackage{psfrag}
\usepackage{amsmath,amssymb}
\usepackage{mathrsfs}

\usepackage{bm}
\usepackage{latexsym}
\usepackage{mathrsfs}
\usepackage[T1]{fontenc}

\newcommand{\be}{\begin{equation}}
\newcommand{\ee}{\end{equation}}
\newcommand{\ba}{\begin{eqnarray}}
\newcommand{\ea}{\end{eqnarray}}
\newcommand{\ban}{\begin{eqnarray*}}
\newcommand{\ean}{\end{eqnarray*}}

\begin{document}

\title{Negative specific heat of black-holes from Fluid-Gravity Correspondence}

\author{Swastik Bhattacharya}\email{swastik@iisertvm.ac.in}

\author{S. Shankaranarayanan} \email{shanki@iisertvm.ac.in}
\affiliation{School of Physics, Indian Institute of Science Education
  and Research Thiruvananthapuram (IISER-TVM), Trivandrum 695016,
  India}

\begin{abstract}
  Black-holes in asymptotically flat space-times have negative specific
  heat --- they get hotter as they loose energy.  A clear statistical
  mechanical understanding of this has remained a challenge. In this
  work, we address this issue using fluid-gravity correspondence which
  aims to associate fluid degrees of freedom to the horizon. Using
  linear response theory and the teleological nature of event horizon,
  we show explicitly that the fluctuations of the horizon-fluid lead
  to negative specific heat for  Schwarzschild black Hole. We also 
  point out how the specific heat can be positive for Kerr-Newman or AdS 
  black holes. Our approach constitutes an important
  advance as it allows us to apply canonical ensemble approach to
  study thermodynamics of asymptotically flat black-hole space-times.
\end{abstract}

\maketitle

\section{ Introduction} Gravity is universally attractive and a
self-gravitating system is unstable to collapse. In Newtonian theory,
it is known that a non-relativistic ideal homogeneous fluid is
unstable to long wavelength density perturbations~\cite{MTW}.
For a gravitating object in a thermal bath, thermal fluctuations will
drive the object momentarily hotter or colder than the bath; since,
gravity is not screened, gravitational object will evolve in whichever
direction it started out with~\cite{LyndenBell}. This is referred to
as negative specific heat paradox~\cite{LyndenBell}.

Black-holes are thermodynamical objects where quantum effects induce
additional instabilities~\cite{BH-Thermo,Hawking-1975}.  Broadly, the
issues in black-hole thermodynamics can be classified into two
categories based on whether or not they can be addressed within the
realm of known Physics. It is widely regarded that understanding of
black-hole entropy or resolution of information paradox require new
Physics
\cite{BHentropy}. However, the issue of negative
specific heat of black-holes and its statistical mechanical
description~\cite{StatmechBH}, may not require new Physics and, if not
resolved, will persist even in quantum gravity.
In this work, we provide a physical understanding of the negativity of
the specific heat and armed with this go on to construct canonical
ensemble description for black-holes.

We use Fluid-Gravity correspondence {\sl as described by Damour} to understand physical origin of
negative specific heat for
black-holes~\cite{FluidGrav,Membrane,FluidGrav2}. Since Damour's 
calculation 30 years ago, attempts have been made to use it to gain new physical
insight from the Fluid-Gravity correspondence. However, there has not been any 
progress in understanding how to go beyond the thermodynamic level in this approach. Recently, many
interesting features of black-holes have arisen since the formal 
relation between the equations of gravity (near the black-hole
horizon) and equations of fluid dynamics were 
found~\cite{FluidGrav,Membrane,FluidGrav2}. More importantly, 
Fluid-Gravity correspondence allows the possibility to connect
macroscopic and microscopic physics through the study of the
statistical properties of the fluid on the black-hole horizon. Current
authors have shown that horizon-fluid is of physical interest, and an
effective theory can be written describing this fluid as a
condensate~\cite{Skakala:2014eba,Bhattacharya:2015yga,Bhattacharya:2015qkt,Bhattacharya:2014xma,Lopez:2015dlu}.
We have shown that Bekenstein-Hawking
entropy\cite{Bhattacharya:2014xma} and Langevin equation for the area
expansion (of 2-D fluid) of homogeneous horizon-fluid is given by the
Raychaudhury equation~\cite{Bhattacharya:2015yga}.

It is well known that fluids can be described by two sets of
parameters, namely susceptibilities and transport coefficients. While
the first set of parameters correspond to changes in local variables;
other set involves fluxes of thermodynamic
quantities~\cite{Kadanoff,Kubo,Zwanzig}.  Obviously, these parameters
can not be determined within fluid mechanics, however, these can be
derived using the theory of fluctuations that relate
susceptibility/transport coefficient to autocorrelation function of a
dynamical variable~\cite{Kadanoff,Kubo,Zwanzig}.  In this paper, we
show explicitly that the fluctuations of the horizon-fluid lead to
negative specific heat (thermodynamic derivative). Specific heat is a 
generalized susceptibility or Thermodynamic 
derivative that quantifies the amount of the change of an extensive quantity (like entropy or internal energy) 
under the influence of an intensive quantity (like temperature).  

We consider fluctuations from the  equilibrium value of the total 
entropy of the horizon-fluid. The change in the total entropy of the horizon-fluid
  is proportional to the change in the temperature of the fluid and
  this proportionality constant is the susceptibility. This method of
calculating susceptibility has its advantages. First, the theory of
Fluctuations is independent of any particular model of quantum
gravity.
Second, using transport theory, the authors have shown that the
teleological nature of the horizon-fluid is responsible for the
negative bulk viscosity of the fluid~\cite{Bhattacharya:2015qkt}. The
teleological nature refers to the fact that the black-hole area starts
increasing long before matter-energy falls in it and stops as the
infalling matter-energy crosses the event horizon. This can be viewed
as the anti-causal response for globally defined
black-holes~\cite{MTW,FluidGrav,Membrane,Bhattacharya:2015qkt}. 
  This is extended here to the case of susceptibilities to show that the 
  Horizon-Fluid and hence black-holes have negative specific heat. The procedure 
  we shall follow is to first define a suitable dynamic susceptibility for the specific 
  heat of the Horizon-Fluid. The static limit of this quantity is shown to be negative 
  and hence, the specific heat is negative.

The rest of the paper is organised as follows.  In the next
  section, we identify the response function corresponding to the
  specific heat of the horizon-Fluid. In the third section, we explicitly 
  evaluate the dynamic susceptibility an then show that the specific heat 
  is negative. In section four, we briefly discuss how some black holes(e.g. rotating charged ones) 
  can exhibit positive specific heat. In the final section, the implications of the results obtained here are briefly discussed.

\section{Identification of the Response Function to determine Specific  heat}
Specific heat is given by,
\begin{equation}
 C = T \, \frac{\partial S}{\partial T}, \label{spht}
\end{equation}
where, $S$ is the entropy and $T$ is the temperature. Since $T$ is positive
definite, the sign of the specific heat is determined by the change of
the entropy of the horizon-fluid due to the change in temperature.
Defining susceptibility as  $\chi_{_T} \equiv \frac{\partial
  S}{\partial T}$~\cite{Kubo}; within linear response theory, the change in the
entropy under the influence of the external influence ($\delta T$) is
given by,
\begin{equation}
  \delta S = \chi_{_T} \,  \delta T  \,. 
  \label{LRchi}
\end{equation}

Typically it is convenient to understand the processes in the frequency $(\omega)$ space and one can define a 
corresponding susceptibility as $\chi_c(\omega)$. In general, this susceptibility, 
$\chi_c$ is complex (See, for instance,  Ref.~\cite{Kadanoff}), i. e. 
\begin{equation}
 \chi_c(\omega+i\epsilon)= \chi'(\omega)+ i\chi''(\omega). \label{ReIm}
\end{equation}
where the imaginary part physically corresponds to absorption. In the static 
limit, there is no absorption and hence only the real part of $\chi_c$ contributes to $\chi_T$.
Physically, the static part corresponds to the quasi-static limit (at long timescales) of the process
with the system always close to equilibrium. The static part of the susceptibility 
corresponds to the specific heat of the fluid. The static susceptibility can be found 
from the dynamic susceptibility using analytic continuation. 

Before we proceed with the evaluation of $\chi_{_T}$, it is useful for the purpose of clarification
to compare and contrast $\chi_{_T}$ with the more familiar magnetic
susceptibility ($\chi_{_M}$). $\chi_M$ satisfies the relation,
$\delta\mathbf{M}= \chi_{_M} \, \delta\mathbf{ H}$, where, $\delta
\mathbf{H}$ is the external influence. In the theory of fluctuations,
$\chi_M$ is the static limit of response function
\cite{Kadanoff,Kubo,Zwanzig}. In the same manner, $\chi_{_T}$ is the
static limit of the response function $\chi''(\omega)$ given by:
\begin{equation}
 \chi_{_T}  = \int\frac{d\omega}{\pi}\frac{\chi''(\omega)}{\omega}. 
\label{eq:StaticChi}
 \end{equation}
\begin{figure}
\begin{center}
  \includegraphics[width=0.25\textwidth]{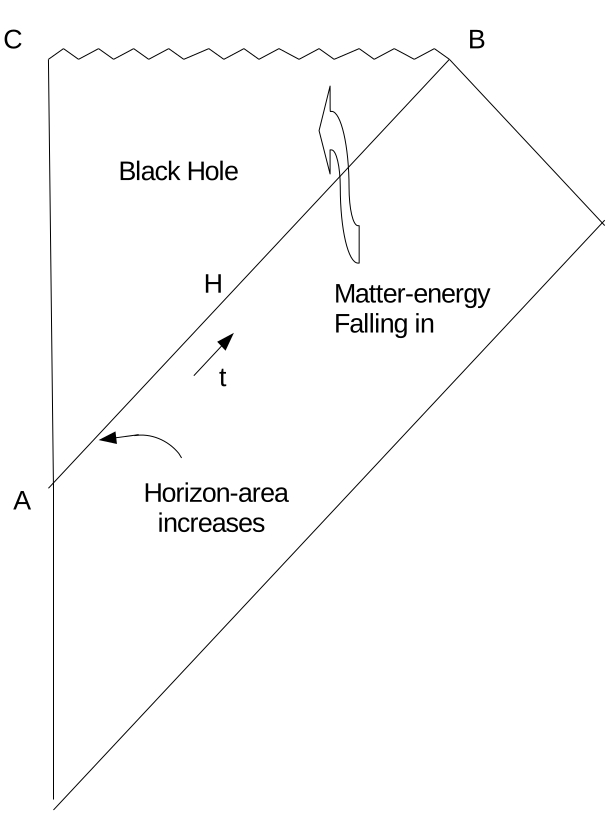}
 \end{center}
 \caption{Penrose diagram of a black-hole space-time. The response of
   a black-hole event horizon is teleological \cite{Membrane}.}
 \end{figure}
 However, there are two crucial differences between $\chi_M$ and
 $\chi_T$:
\begin{enumerate}
\item It is possible to evaluate $\chi_M$ from fundamental principle
  by writing down the Hamiltonian taking into account the quantum
  effects.  In our case, since we do not have complete information
  about the evolution dynamics, the time evolution of $\delta S$ is
  governed by a phenomenological generalized Langevin
  equation~\cite{Bhattacharya:2015yga}.  However, still the excess
  entropy density and entropy current satisfy continuity equation~[See
  Appendix for details].
\item While the response in the case of magnetic system is causal,
  here it is anti-causal. This is because, for the event horizon, the
  response to any external influence is anti-causal.  In particular,
  if matter-energy falls through the event horizon, the area of the
  event horizon increases till the matter-energy passes through the
  horizon~\cite{Membrane}. This is physical as the event horizon 
  of a black hole is defined globally in the presence of the future light-like infinity \cite{Membrane}.

  Due to this unusual property of the horizon, the horizon-fluid also
  exhibits anti-causal response  i.e. the response of the horizon takes 
  place before the external influence occurs~\cite{Membrane}.
  This is referred to as the teleological nature of horizon (See Fig
  1). For canonical ensemble, the specific heat is proportional to the energy fluctuations squared i. e.,
$$C =  \frac{(\Delta E)^2}{T^2} \, . $$
 \vspace*{-20pt}
 
 While for a causal process, it is positive; as we will show below, for an anti-causal fluid, the specific heat turns out to be negative.
\end{enumerate}

\section{Teleological boundary condition and Negative specific heat}

Thermal ensemble for small fluctuations away from the equilibrium is
characterised by the entropy change due to
fluctuations\cite{Bhattacharya:2015yga,Bhattacharya:2015qkt},
\begin{equation}
 \Delta S = \int \delta S = - \frac{1}{4 A} \int \delta A \, d(\delta A) =  -\frac{1}{8} \frac{(\delta A)^2}{A}. \label{DeltaSfl}
 \end{equation}
 It is important to note that $\Delta S$ is negative because the system is not in equilibrium and hence 
 the entropy of the system is less than that in the equilibrium state. 
In the rest of the section we evaluate explicitly the susceptibility
 ($\chi_T$), and show that $\chi_T$, and hence the specific heat,
 is negative. The procedure we adopt is the following:  
 \begin{enumerate}
  \item   First, we express  the dynamical susceptibility as an integral 
  involving autocorrelation function of the entropy density fluctuations. 
  \item We then evaluate the  static susceptibility from the dynamic susceptibility using 
  analytic continuation. Using the teleological boundary condition we show that the specific heat 
  for the Horizon-Fluid is negative.
 \end{enumerate}
 
 \subsection{Dynamic Susceptibility from Autocorrelation of entropy
   fluctuations}
   
For a linear response, the ensemble average of
 the fluctuations in entropy ($\delta S$) described by Eq.~(\ref{LRchi})
  in the frequency domain is given by:
\begin{equation}
  \langle\delta S[\omega]\rangle = 
  \int_0^\infty\chi''(\omega)e^{-i\omega t}\delta T(t)dt, \label{chit}
\end{equation}
where, $\chi''(\omega)$ is dynamic susceptibility and is given by
\cite{Kadanoff,Kubo,Zwanzig},
\begin{equation}
  \chi''(\omega)= -\frac{1}{8\pi\langle\delta S^2(0)\rangle}\int_{-\infty}^{\infty}\langle \delta S(0)\delta S(t)\rangle e^{i\omega t} dt. \label{chiab}
\end{equation}

The linear response of the system can be written as,
{\small
\begin{eqnarray}
  \langle\delta S(t)\rangle = -\int_0^\infty \!\!\! d\omega \int_0^{\infty} \!\!\! dt' dt'' \frac{\langle \delta
  S(0)\delta S(t'')\rangle} {2\langle\delta
    S^2(0)\rangle} \delta T(t')  e^{i\omega(t-t')}e^{i\omega t''} \!\!
\end{eqnarray}
}
which leads to
\begin{align}
\label{deltaS2}
 \langle\delta S(t)\rangle= -\frac{1}{2\langle\delta
   S^2(0)\rangle}\int_0^\infty\int_0^{\infty}& \langle \delta
 S(0)\delta S(t'')\rangle  \\ & \!\!\!\!\!\!\!\!\!\!\!\!\!\!\!\!\!\! \delta
 T(t')\delta(t-t'+t'')T(t')dt'dt'' \nonumber 
\end{align}
Integration of the R.H.S. of the above equation leads to a theta
function that decides whether the response is causal or
anti-causal. Taking into account, the time translation invariance of the
auto-correlation function, we get,
\begin{equation}
 \langle\delta S(t)\rangle= \frac{1}{2\langle\delta
   S^2(0)\rangle}\int_t^\infty \langle \delta S(t)\delta S(t')\rangle
 \delta T(t') dt', \label{ACResponse1}
\end{equation}
where, $t'>t$. It is to be noted here that \eqref{ACResponse1}
explicitly shows the anti-causal nature of the response function. The
linear response can now be expressed as,
{\small
\begin{equation}
 \langle\delta S(t)\rangle= -\frac{1}{2\pi} Im\int_t^\infty\int
 d\omega\chi''(\omega)\delta T(t')e^{-i\omega(t-t')}dt'
 \label{eq:dynamicSus}
\end{equation}
}
Having obtained the dynamic susceptibility, our next step is to 
obtain the static susceptibility and use Eq. (\ref{LRchi}) to evaluate 
the specific heat.

\subsection{Evaluation of specific heat}

Since we want to look at the static limit of the response, we take
$\delta T(t)=\delta T$, i.e. a constant.  From Eq.~(\ref{eq:dynamicSus}), we get
\begin{equation}
 \langle\delta S(t)\rangle= \delta T\int\frac{dw}{w}\chi''(\omega). \label{Responsechi}
\end{equation}

As mentioned above, we are interested in evaluating the static limit of the susceptibility. Following  \eqref{ReIm} and using 
Kramers-Kronig  relations\cite{Kadanoff}, leads to,
\begin{equation}
 \chi'(\omega)= - P
 \int\frac{d\omega'}{\pi}\frac{\chi''(\omega')}{\omega'-\omega}. \label{KKR}
\end{equation}

From \eqref{KKR} and \eqref{Responsechi}, we get, 
\begin{equation}
 \langle\delta S(0)\rangle= \chi'(0) \delta T. \label{LR1}
\end{equation}
Since the response at time zero is a response to an adiabatically
applied disturbance, $\chi'(0)$ is the static susceptibility, i. e.,  
$\chi'(0)= \chi_\tau$.  Eq. \eqref{LR1} leads to $\langle\delta S(0)\rangle= \chi_\tau \delta T$. 
This leads to \eqref{eq:StaticChi}, the expression for the static
  susceptibility. 

Anti-causal response ($\delta S(t)$) (is proportional to the
inhomogeneous part of the density that) oscillate slowly in time, has
a thin spread $\Delta \omega$ and is given by,
\begin{equation}
  \delta S(t)= \theta(-t)\int_{\Delta \omega}e^{i\omega t} d\omega. \label{Somega}
\end{equation}
Putting these back in \eqref{chiab}, one gets, 
\begin{equation}
  \chi''(\omega)= -\frac{1}{8\pi}Im\int_{\Delta \omega}d\omega\int_{-\infty}^{\infty}\theta(-t)e^{i\omega't}e^{i\omega t}dt  = -\frac{1}{8 \pi} \, .
\label{TCchi}
\end{equation}
The absorptive part of the dynamic susceptibility comes out to be
negative due to the teleological boundary condition for the
Horizon-Fluid. It is straightforward to check that $\chi''$ is
positive for a causal response, $\delta_C S(t)= \delta S(t)=
\theta(t)\int_{\Delta \omega}e^{i\omega t} d\omega$, hence $\chi$ is
positive.

Substituting the above value of $\chi''(\omega)$ in Eq. ~\eqref{eq:StaticChi}, 
static susceptibility is given by:
\begin{equation}
 \chi_{_T} = -\frac{1}{8\pi^2}\int \frac{d\omega}{\omega}. \label{ChiS}
\end{equation}
\eqref{ChiS} shows clearly that $\chi_{_T}$ is negative. Hence, the
horizon-Fluid specific heat is negative.
It is important to note that the integral is well-defined with
infrared (corresponding to the horizon size) and ultraviolet cutoffs.

\section{Understanding positive Specific Heat of black-Holes from fluctuation dissipation}

In the previous section, we have shown explicitly that within the transport theory it is possible to obtain 
a fundamental understanding of the negative specific heat of a Schwarzschild 
black hole. A natural question that arises now is whether the same analysis can provide an explanation for the 
positive specific heat for certain ranges of 
parameters, for charged black holes with/without spin \cite{Davies} and for asymptotic AdS black-holes. In the rest of this 
section, we provide arguments to show how the 
analysis changes for these cases and how positive values of specific heat may come about. 

\subsection{Asymptotically flat black holes} 

In order to extend the analysis to general stationary black-hole space-times, let us relook at the 
calculation for such a black-hole from the point of view of the Raychaudhury equation of a null congruence 
on the event horizon. Let us first write down the Raychaudhury equation for a quasi-stationary charged, spinning black hole to 
emphasize the same general form all of them can be expressed in. 
\begin{equation}
 -\frac{d\theta^H}{dt}+ g_H\theta^H= [8\pi(\mathcal{I}^H+4 \pi |\mathbf{K}|^2)+ \sigma_{AB}\sigma^{AB}]-\frac{1}{2}{\theta^H}^2, \label{Thetagen}
\end{equation}
where, $\theta_H$ is the expansion scalar, $\mathcal{I}^H$ denotes the 
energy flux through the horizon, $g_H$ is the surface gravity on the horizon, $\mathbf{K}$ is 
the surface current on the black-hole surface and $\sigma_{AB}$ is the shear term.
Though \eqref{Thetagen} is exact, in what follows, we shall also assume $\theta_H$ to be small. Later, we  point how 
a charged and spinning black hole would exhibit different response  from one which is neutral and non-rotating. 

Since the above equation is non-linear in $\theta^H$, it is difficult to perform a stability analysis 
for the black hole horizon starting from it.
Rewriting in terms of  $\eta= \sqrt{\mathcal{A}}$, i. e. 
\begin{equation}
 \theta= \frac{d}{dt}(\ln{\mathcal{A}}) \, ; \frac{d\theta}{dt}= \frac{1}{\mathcal{A}}\frac{d^2}{dt^2}(\mathcal{A})-\frac{1}{(\mathcal{A})^2}(\frac{d\mathcal{A}}{dt})^2. \label{ThetaA}
\end{equation}
where $\mathcal{A}$ is the area of cross-section of the null congruence on the event horizon. 
Rewriting (\ref{Thetagen})\footnote{Writing $\mathcal{A}= \mathcal{A}_0+ \delta\mathcal{A}$, where, $\delta\mathcal{A}$ is the change in this area 
over some constant base value $\mathcal{A}_0$. Similarly, $\eta$ can also be expressed as $\eta=\eta_0+\delta\eta$.}  and 
using  $g=8\pi T$\cite{Bhattacharya:2015yga}, we can rewrite \eqref{Thetagen} as
\begin{equation}
 -\ddot{\eta}+2\pi T\dot{\eta}\simeq \mathcal{S}, \label{eta}
\end{equation}
where, 
\begin{equation}
 \mathcal{S}= \frac{\sqrt{\mathcal{A}}}{2}[8\pi (\mathcal{I}^H+ 4\pi |\mathbf{K}|^2)+ \sigma_{AB}\sigma^{AB}]. \label{etaSource}
\end{equation}
It is important to note that Eq. \eqref{eta} holds only up to second order in $\delta\eta$ and it applicable to any general, stationary black-hole. 
$\mathcal{S}$ can be viewed as a source term. 
As we know, the black hole mass or the energy and the entropy of the horizon-fluid increases 
when matter-energy passes through the horizon~\cite{Bhattacharya:2015yga}. Hence, the matter-energy flux
across the horizon, $\mathcal{I}^H$ acts as a source term. 
But it is not the only term that drives the evolution of $\eta$. There are two other terms as well, 
one describing the effects of the shear and is proportional to $\sigma_{AB}\sigma^{AB}$; the 
other the effect of an electric current on the black hole horizon and is proportional to $|\mathbf{K}|^2$.
As we shall see, whether the specific heat is positive or not depends on which of these three terms 
is dominant.

\subsubsection{Schwarzschild black-holes}

Let us now focus on the Schwarzschild black hole which is the final stationary state of a 
nonrotating neutral black hole. Eq. (\ref{eta}) becomes  
\begin{equation}
 -\ddot{\eta}+2\pi T\dot{\eta}\simeq 4\pi\sqrt{\mathcal{A}}\mathcal{I}^H, \label{etaS}
\end{equation}
where, $\mathbf{K}=0$ and $\sigma_{AB}=0$.

From the LHS of \eqref{etaS}, one sees that $\eta$ would increase exponentially with time. Demanding that the horizon exists 
in the future necessitates one to impose future boundary condition on $\eta_H$ \cite{Membrane} instead of the initial 
boundary condition. (Usually, one imposes the future boundary condition on $\theta^H$ but it amounts to the same thing.)  
Otherwise $\eta$ would increase exponentially and destroy the black-hole. Here we recall that we have assumed that $\eta$ to be small, which 
means we have to ensure through the boundary condition that it never grows to a large value. 
Thus the teleological boundary 
condition can be viewed as a condition for the stability of the black-hole event horizon. From the fluid perspective, 
this can be viewed as the condition that small external influences are not going to drive the system far from 
equilibrium \cite{Bhattacharya:2015yga}. As has already been demonstrated, 
this can be attributed to the negative specific heat for the Schwarzschild black-hole.

\subsubsection{Charged black-holes}

Now let us look at the charged black-hole.  Eq. (\ref{eta}) becomes  
\begin{equation}
 -\ddot{\eta}+2\pi T\dot{\eta}\simeq \frac{\sqrt{\mathcal{A}}}{2}8\pi (\mathcal{I}^H+ 4\pi |\mathbf{K}|^2)
 \label{SourceRN}
\end{equation}
In the  equilibrium limit, the system settles down to a Reisner-Norstrom black-hole. The electrical potential 
$\phi$ is constant on the surface of a black hole in that state. This implies that no current flows at equilibrium. 
In the fluid picture, although the fluid is charged, there is no current as the velocity of the horizon-fluid for 
a Reisner-Norstrom black-hole is zero.

Let us consider a process in which the black-hole mass increases, however, the black-hole charge remains a 
constant. The specific heat $C_Q$ is given by $C_Q= T(\frac{\partial S}{\partial T})_Q$. The electrostatic potential, 
$\phi$ on the black-hole surface  would not be constant if the black-hole is away from its equilibrium state. Thus, 
for a charged black-hole undergoing a change in its mass, $\phi$ vary on the surface. However, 
the black-hole horizon has a conductivity given by $\frac{1}{4\pi}$\cite{FluidGrav}, \cite{Membrane}. So 
the inhomogeneous potential would generate an electric field on the black-hole surface. This would produce 
a current in the fluid that would quickly neutralize the electric field. Such currents 
would be exponentially damped in time. Now, for a stationary charged black-hole, $\phi\propto Q$. This implies that 
$\mathbf{E}=-\nabla\phi\propto Q$. Because Ohm's law is applicable on the black-hole horizon\cite{FluidGrav}, 
we have $|\mathbf{K}|\propto Q$. 

Thus, the change from the case of a non-rotating neutral black-hole is the introduction of a second term in 
\eqref{SourceRN} as a source along with the term $\mathcal{I}^H$. The two source terms, $\mathcal{I}^H$ and 
$|\mathbf{K}|^2$ are different in nature in the sense that individually they would generate anti-causal and 
causal responses in the horizon-fluid. { To see how this comes about, first we note that the matter-energy 
flux, $\mathcal{I}^H$ that passes through the horizon does not get damped in time(Killing time).} However, 
as we discussed above, the surface current $\mathbf{K}$ for the process we  consider would be exponentially 
damped in time. 

To bring out the difference in the response explicitly, we consider what would have 
happened had $\mathcal{I}^H=0$ but $\mathbf{K}\neq 0$. We emphasize that this example is chosen for 
demonstrative purpose only and does not correspond to the situation we are considering. In the process we 
have been considering, if $\mathcal{I}^H=0$, then $\mathbf{K}=0$ also. Nonetheless, let us proceed with this 
example and write down the evolution equation for $\eta$ in this case, which, from \eqref{eta}, we see 
then would have the form,
\begin{equation}
 -\ddot{\eta}+ 2\pi T \dot{\eta}\simeq 16\pi^2 \sqrt{\mathcal{A}}|\mathbf{K}|^2 .\label{etaCurrent}
\end{equation}
As we have already argued that $\mathbf{K}$ has a damping factor of the form $e^{-\varsigma t}$, it is 
seen that subject to initial boundary conditions, equation 
\eqref{etaCurrent} has solutions, where 
$\eta$ remains finite throughout, if $\varsigma>2\pi T$. This is in clear contrast to \eqref{etaS}, where one 
needed to impose a future boundary condition to keep $\theta$ finite. Thus the surface current $\mathbf{K}$ acting 
as a source can generate a causal response whereas, for matter-energy flux as the source term, the response is necessarily 
anti-causal. 

It is clear that for the process, we have been considering, both the terms on the right hand side of 
\eqref{SourceRN} would compete with each other and depending on the relative magnitude of these, it would be 
decided whether the response would be causal or anti-causal. Since $|\mathbf{K}|^2\propto Q^2$, it is 
natural that when $Q$ is small, the response of the horizon would mainly be driven by the 
matter-energy flux and would be anti-causal, hence giving rise to negative specific heat. However, as $Q$ 
increases, there might come a point where the surface current that is generated becomes the main external 
driving force, thus making the response causal. In this case, the specific heat of a charged black-hole 
would become positive. To see how this comes about, one has to consider the Transport theory for a 
charged black-hole in detail, something that is out of the scope of this work. Also, one needs to be careful 
while considering the fluctuations near the point, where the transition from negative to positive specific 
heat takes place, since there is some evidence that a phase transition takes place at this 
point \cite{Davies}. 

\subsubsection{Rotating black-holes}

Moving on to a spinning black-hole that is neutral, we see that one can argue in a similar manner. 
In this case, the specific heat is given by $C_J= T(\frac{\partial S}{\partial T})_J$. The process that 
we have to consider here is an increase in the mass of the black-hole while keeping its total angular momentum $J$ 
fixed. The equilibrium state for such a spinning black-hole is the Kerr black-hole, which is stationary. 
The Smarr relation for the Kerr black-hole is given by, 
\begin{equation}
 \frac{\kappa A}{4\pi}= M- 2\Omega_H J, \label{KSmarr}
\end{equation}
where, $\Omega_H$ is the angular velocity of the black-hole. Comparing this with the Smarr relation for a 
Reisner-Norstrom black-hole, we see that the structure is similar once $Q$ is replaced by $J$ and $\phi$ by 
$\Omega_H$. The argument given in the case of a charged black-hole can be constructed for a spinning 
black-hole based on this similarity. Away from equilibrium, $\Omega_H$ should change from its equilibrium 
value and would vary from point to point on the black Hole surface if we take into account the 
inhomogeneities. 

However, for a spinning black-hole, 
\begin{equation}
 \mathbf{v}\equiv \Omega_H \frac{\partial}{\partial\phi}, \label{vomega}
\end{equation}
i.e. the angular velocity is the velocity of the horizon-fluid\cite{FluidGrav}. Hence such variation in the 
velocity of the horizon-fluid would generate a shear term, $\sigma_{AB}$ in the fluid. Now we recall that 
the shear is generated by gravitational waves\cite{Membrane}. Moreover, the gravitational 
wave modes that generate shear on the black-hole horizon are damped and decay exponentially in 
time\cite{Membrane}. If we now consider the evolution equation for the shear term on the null congruences 
on the black-hole event horizon\cite{Membrane}, it is seen that the shear term itself would die down 
exponentially in time. Finally before looking into the transport process, we recall that $\Omega_H\propto J$ 
for a Kerr black-hole. For small deviations from the Kerr black-hole, this proportionality is expected to 
hold. 

Now let us look at \eqref{eta} on the event horizon of a spinning black-hole as in the previous cases, 
\begin{equation}
 -\ddot{\eta}+ 2\pi T \dot{\eta}\simeq \frac{\sqrt{\mathcal{A}}}{2}(8\pi\mathcal{I}^H+ \sigma_{AB}\sigma^{AB}), \label{etaK}
\end{equation}
where, the surface current, $\mathbf{K}=0$.
Now one can proceed exactly as before and argue how the positive specific heat of a spinning black-hole might 
come about. For a spinning, charged black hole, similar considerations will again apply. 

\subsection{Black-holes in AdS Background}

It is well known that AdS black-holes can have positive specific heat for certain ranges of the value of the 
Cosmological constant(negative $\Lambda$). It is not within the scope of this work to show how this occurs. 
However we can point out a possible direction along which the explanation may lie. The main difference between 
a Schwarzschild and an  AdS-background black hole is the presence of a negative cosmological 
constant($-\Lambda$) in the latter case. This, however, does not make any contribution in the equation for 
the horizon-fluid as such a fluid is defined on a null hypersurface \cite{Bethan}. Hence, black-holes in AdS background 
should be considered directly as a thermodynamic system and later fluctuation theory needs to be applied.
A possible way to proceed is to treat $\Lambda$ as a thermodynamic variable. Then one can define another
thermodynamic variable conjugate to it (See, for instance, \cite{AdS1}, \cite{AdS2}\cite{AdS3}). Now one has to apply the
Fluctuation-Dissipation 
theorem to this system and try to determine its specific heat. It is clear however that the analysis is 
going to be quite different from the one done for a Schwarzschild black hole and the argument described 
here leading to the negative specific heat of a black hole would not be applicable to this case.

\subsection{Dynamical and future outer trapped horizons}

In Refs. \cite{Hayward,DH}, the authors have explicitly shown how to define dynamical horizon or the future outer trapped horizon 
and also have described the construction of horizon-fluid. Unlike the event horizons, these horizons are defined locally and 
they are not Killing horizons. 

In the context of horizon-fluid, one major difference between the dynamical horizons and event horizons is  that the congruence on such 
a dynamical horizon has a non-zero twist $\mathbf{\omega}$\cite{DH}. This changes the horizon-fluid equation 
for such a horizon in a qualitative manner as the change in the  energy of such a fluid is 
not given only by the matter-energy flux through the horizon but there is also an additional heat flux 
in the form, $\mathbf{\mathcal{D}}.(\frac{\sqrt{h}}{4\pi}\mathbf{\omega})$, i.e. proportional to the 
divergence of the 2-form $\mathbf{\omega}$.   Thus the total heat flux 
across the dynamical horizon is different from that for the event horizon of Black hole in a 
dynamical spacetime even if the Black hole mass increases by the same amount\cite{DH}. From the 
previous analyses, this suggests that the specific heat can be different for the horizon-fluids for different horizons.

We find this to be the case when we consider the Theory of Fluctuations for these 
horizons. Since, the dynamic horizon or the Future Outer Trapped Horizon is 
defined locally, one need not impose a future boundary condition for the evolution of 
$\theta$ of a null congruence on the horizon. However, 
imposing an initial condition means that the response of the horizon-fluid corresponding to  
a Dynamic Horizon or a Future Outer Trapped Horizon is causal in nature as opposite to the 
case of Schwarzschild black holes, where the response was anti-causal. This 
makes the specific heat of the horizon-fluid corresponding to the dynamic horizon of a black hole 
positive. 

\section{Discussion} 
In late 1960's and early 1970's, the mathematical analogy between
black holes and ordinary thermodynamics was
established~\cite{BH-Thermo}. However, only after Hawking's famous
discovery of the evaporation of black-holes~\cite{Hawking-1975}, it
was realised that the pairs of analogues between black-holes and
thermodynamics are indeed physically similar. Likewise, mathematical
similarity between equations of General relativity near the black-hole
horizon and fluids was known for a long time~\cite{FluidGrav,Membrane}.
While it is treated as a mathematical curiosity, we have shown that
the horizon-fluid itself is of physical interest. Using
fluctuation-dissipation theorem~\cite{Kadanoff,Kubo,Zwanzig}, we have
shown explicitly that it is possible to derive the transport
coefficients like bulk viscosity or Thermodynamic derivatives.

Till now, there has been no statistical mechanical understanding as to
why asymptotically flat space-time black-holes have negative specific
heat. In this work --- using fluctuation-dissipation theorem and that
the black-hole horizon has anti-causal response --- we have provided a
statistical mechanical understanding of a long-standing problem in
black-hole Physics. An important consequence of our approach is that
it allows black-holes to be described in canonical ensemble
picture. In canonical ensemble, specific heat is related to the
mean-squared fluctuations of energy \cite{Huang},
\begin{equation}
C = \frac{1}{T^2}\left[ \langle \Delta E^2 \rangle \right] \, . 
\end{equation}
Since RHS is positive definite, it is not straightforward to describe
systems with negative specific heat~ \cite{Huang}.  The inference
above, makes an implicit assumption about the causal response. As we
have shown above for systems that are anti-causal, specific heat is
negative (See also Evans and Searles \cite{1996-Evans.Searles-PRE}). Thus, our analysis explicitly shows
that black-holes can be treated within the canonical ensemble in
statistical mechanics(provided future boundary condition is used);
probability for such a system to be in a state
is $P_{\Delta E}\propto\exp{\frac{-\Delta E}{T}}$. Summing over all
possible states, one can write down a partition function for the
system as:
\begin{equation}
 Z  = \sum_{\mbox{all states}} \exp \left[ \frac{-\Delta E}{T} \right]
\end{equation}

To keep the physics transparent, we have restricted our analysis to
Schwarzschild black-hole. However, fluctuation-dissipation analysis as applied here
can be extended to the other black-holes, like Reisner-Norstrom
black-hole, Kerr black-hole. Mapping the 
fluid equations to Raychaudhury equation, we have argued that it 
is possible to explain positive specific heat for certain ranges of parameters 
for Reisner-Norstrom and Kerr black-holes. It is interesting to perform 
the complete analysis for these general cases. This is currently under progress.

 It is pertinent here to bring up the issue as to where our analysis stands vis-a vis the negative 
specific heat observed for self-gravitating bodies in Newtonian Gravity. Obviously, a future 
boundary condition is not required for solving the equation for heat transport in such a case and our 
formalism does not apply to those cases. As 
mentioned in the Introduction, one can construct a microcanonical ensemble for such a system and 
calculate the specific heat which turns out to be negative. Unfortunately, things are not so straightforward 
when we consider black-holes. This is because Black holes differ sharply from any other system composed of 
matter. First, the entropy of the black-hole is not extensive. Secondly, for a self-gravitating system 
in Newtonian theory, we know what the underlying degrees of freedom are. This is not the case for Black 
Holes and there has been lot of debate regarding the relevant degrees of freedom. This 
greatly restricts our ability to say anything concrete about the negativity of Black hole specific heat 
by constructing microcanonical ensembles. The only thing on which there seems to be some amount of agreement 
in the literature is that these underlying degrees of freedom has to be related to the space-time geometry in some way. This is 
qualitatively different from the case of self-gravitating body in Newtonian gravity, where one looks only 
at the matter degrees of freedom. Hence the explanation for the negativity of the specif heat for these 
two entirely different objects may turn out to be quite different. Under this 
circumstance, it is a step forward in understanding the specific heat of black-holes if an 
explanation can be provided for its negativity based only on the general properties of fluctuations and the 
teleological nature of the event horizon. 

Our analysis
is applicable in principle to asymptotically flat and AdS black-hole space-times.
Until now, such a relation has been obtained only for black-holes in
AdS background~\cite{FluidGrav2}; attempts to generalize~\cite{KCFT}
has not been so successful.  On the contrary, accumulating evidence
from this study and other
studies~\cite{Bhattacharya:2015yga,Bhattacharya:2015qkt,Bhattacharya:2014xma,Bethan}
suggest that the horizon-fluid fluctuations have deep statistical
mechanical significance and relates to Gravity description. Important
insight can be obtained if one can map the theory of Fluctuations in
AdS black-hole space-times with AdS-CFT, which possibly may provide a
field theoretic description for Gravity.

\section*{Acknowledgments}
The authors thank T. Padmanabhan for comments on an earlier version of the manuscript.
The work is supported by Max Planck-India Partner Group on Gravity and Cosmology. 

\appendix*
\section{Constructing the Continuity Equation}

To find out the susceptibility related to the change in the entropy of
the fluid, we look at the conserved current, i.e. it satisfies a
continuity equation\cite{Kadanoff},\cite{Kubo}. This, as will be seen,
can be interpreted as the excess entropy current and the excess
entropy density of the Horizon Fluid. Here it is important to 
stress the fact that this however does not constitute the total 
entropy current but only the part corresponding to the change in the area of 
the horizon-fluid. The total entropy current must take into account the heat diffusion 
and shear viscosity processes into account. In what follows, we shall
assume that the entropy of the Horizon-Fluid at any point of time is
given by, $S=S_0+\delta S$, where, $S_0$ is the entropy of the system
at the equilibrium state and $\delta S$ is the change in the entropy
from its equilibrium value.

In this case, $N\propto A$ and $A\propto S$, hence $\delta S
\propto\delta N$. Keeping this in mind, we define a physical quantity
called excess number density, $\delta\rho_N$, where,
$\delta\rho_N\propto\delta\rho_S$, the entropy current density and
\begin{equation}
 \int (\delta\rho_N) dA = \delta N. \label{rhoN}
\end{equation}
Here we would like to emphasise that the change in entropy $\delta S$ as the system is away from the equilibrium state 
is negative and so is the change in the total number of excess DOF, $\delta N$. As the 
horizon-fluid's equilibrium state is in the future, $\delta N$ and $\delta \rho_N$ always remain  negative.
The continuity equation for $\delta\rho_N$ is 
\begin{equation}
 \frac{\partial(\delta\rho_N)}{\partial t}+
 \nabla. (\delta\rho_N\mathbf{v})=-\mathcal{S}(\mathbf{r},t). \label{ContE}
\end{equation}
Here $\mathbf{v}$ is the velocity of the two dimensional
horizon-fluid and $\mathcal{S}(\mathbf{r},t)$ is a source term. Typically Eq. \eqref{ContE} is 
written by taking the source term to be zero. However, in this case, the source term is non-zero and is responsible for the 
matter-energy infusion into the horizon-fluid. Since the process is anti-causal, we have taken $\mathcal{S}>0$ physically corresponding to
a sink term, also a consequence of having an equilibrium state in the future. 

As a consequence of the conservation of the excess number of DOF, we have
\begin{equation}
 \frac{d\rho}{dt}-\mathcal{S}= \frac{\partial \rho}{\partial t}+\mathbf{v}.\nabla\rho \, . 
 \label{drho}
\end{equation}
From \eqref{ContE} and \eqref{drho}, one gets, 
\begin{equation}
 \nabla.\mathbf{v}= -\frac{1}{\rho}\frac{d\rho}{dt}. \label{Thetarho}
\end{equation}
One can express $\delta\rho_N$ as, 
\begin{equation}
\delta\rho_N= \frac{\delta N}{\langle A_0\rangle}+ higher\ order\ terms.
\end{equation}

Keeping terms up to first order, we get from \eqref{Thetarho},
\begin{equation}
 \nabla .\mathbf{v}= \frac{\partial}{\partial t}(\ln\delta N)=
 \frac{\partial}{\partial t}(\ln\delta A). \label{NCEN}
\end{equation}
Because of the homogeneity of the system, \eqref{NCEN} implies
\begin{equation}
 \nabla .\mathbf{v} = \frac{\partial}{\partial t}(\ln{A})= \theta, \label{thetaCE}
\end{equation}
where, $\theta$ is the volume expansion coefficient of the null
geodesic congruence on the horizon.  \eqref{thetaCE} is a physically
reasonable requirement and is known to hold for the horizon fluid
\cite{FluidGrav},\cite{Bhattacharya:2015yga}.

Having written down the continuity equation relevant for this
process, we need to show that the change in entropy as the system
fluctuates between the equilibrium and the non-equilibrium states is
given by \eqref{DeltaSfl}. It is important to note that the
Horizon-Fluid is a one parameter system, so the fluctuation away from
the equilibrium state is also characterised by a single parameter,
$\delta A$ or $\delta N$. This means whatever are the variables that
are fluctuating, the change in the Horizon- Fluid entropy would still
be given by \eqref{DeltaSfl}. To verify this, we use \eqref{thetaCE}
to get,
\begin{equation}
 \frac{d}{dt}(\delta S)=  \delta S \frac{d}{dt}(\ln A). \label{Sdot}
\end{equation}
From $\delta S= -\frac{1}{4} \delta A$, up to second order in $\delta
A$, we get,
\begin{equation}
  \delta_t(\delta S)= \delta S(t_2)- \delta S(t_1)= \frac{1}{8}\frac{\delta A^2}{A}, \label{LdeltaS}
\end{equation}
with $t_2>t_1$.  It is important to note that  the change in entropy 
is positive because  $\delta_t(\delta S)$ is the difference in entropy of the fluid
  between $t_2$ and $t_1$. Since the horizon-fluid is moving
  towards the equilibrium state, $\delta S$ is negative and $\delta
  S(t_2)>\delta S(t_1)$. It is to be noted that \eqref{LdeltaS} is
valid only up to second order. This establishes that the process we
are describing is indeed a fluctuation around the equilibrium state of
the system.

\end{document}